\documentclass[10pt,journal,compsoc,oneside,openany]{IEEEtran}
\ifCLASSOPTIONcompsoc
  \usepackage[nocompress]{cite}

  \usepackage{graphicx}
  \usepackage{subfigure}
  \usepackage{booktabs}
  \usepackage{cite}
  \usepackage{pdfpages}
  \usepackage[colorlinks,linkcolor=red]{hyperref}
   \usepackage[center]{caption}
  \usepackage{newpythonhighlight}
  \usepackage{listings}
  \usepackage[numbered, autolinebreaks,useliterate]{mcode}
  \usepackage{color}
  \usepackage{times}
  \usepackage[utf8]{inputenc}
  \usepackage[T1]{fontenc}
  \usepackage{mathrsfs,amsmath}
  \usepackage{amssymb}
  \usepackage{bm}
  \usepackage{longtable}
  \usepackage{algorithmic}
  \usepackage[ruled,linesnumbered]{algorithm2e}
  
\else
  \usepackage{cite}
  \usepackage{xcolor}
  \usepackage{graphicx}
  \usepackage{subfigure}
  \usepackage{booktabs}
  \usepackage{cite}
  \usepackage{pdfpages}
  \usepackage[colorlinks,linkcolor=red]{hyperref}

  \usepackage{newpythonhighlight}
  \usepackage{listings}
  \usepackage[numbered, autolinebreaks,useliterate]{mcode}
  \usepackage{color}
  \usepackage{times}
  \usepackage[utf8]{inputenc}
  \usepackage[T1]{fontenc}
  \usepackage{mathrsfs,amsmath}
  \usepackage{amssymb}
  \usepackage{bm}
  \usepackage{longtable}
  \usepackage{algorithmic}
  \usepackage[ruled,linesnumbered]{algorithm2e}
  
\fi
\ifCLASSINFOpdf
\else
\fi

\begin{document}
%
\title{Reward-Reinforced Generative Adversarial Networks for Multi-agent Systems}
%
%
%
%

\author{Changgang~Zheng,~Shufan~Yang$^*$,~Juan Parra-Ullauri,~Antonio~Garcia-Dominguez,~and~Nelly~Bencomo  
\IEEEcompsocitemizethanks{\IEEEcompsocthanksitem Changgang Zheng is at Department of Engineering Science, University of Oxford, United Kingdom; Shufan Yang ${^*}$ is the corresponding author and currently is School of Computing, Edinburgh Napier University, Edinburgh, United Kingdom and Center of Medical and Industrial Ultrasonics, University of Glasgow, Glasgow, United Kingdom \protect E-mail:s.yang@napier.ac.uk;~Juan Parra,  ~Antonio~Garcia-Dominguez,  and~Nelly~Bencomo are with School of Engineering and Applied Science, Aston University, United Kingdom.
\protect  2021 IEEE.  Personal use of this material is permitted.  Permission from IEEE must be obtained for all other uses, in any current or future media, including reprinting/republishing this material for advertising or promotional purposes, creating new collective works, for resale or redistribution to servers or lists, or reuse of any copyrighted component of this work in other works.}

}

\IEEEtitleabstractindextext{%
\begin{abstract}
Multi-agent systems deliver highly resilient and adaptable solutions for common problems in telecommunications, aerospace, and industrial robotics. However, achieving an optimal global goal remains a persistent obstacle for collaborative multi-agent systems, where learning affects the behaviour of more than one agent. A number of nonlinear function approximation methods have been proposed for solving the Bellman equation, which  
describe a recursive format of an optimal policy. However, how to leverage the value distribution based on reinforcement learning, and how to improve the efficiency and efficacy of such systems remain a challenge. In this work, we developed a reward-reinforced generative adversarial network to represent the distribution of the value function, replacing the approximation of Bellman updates. We demonstrated our method is resilient and outperforms other conventional reinforcement learning methods. This method is also applied to a practical case study: maximising the number of user connections to autonomous airborne base stations in a mobile communication network. Our method maximises the data likelihood using a cost function under which agents have optimal learned behaviours. This reward-reinforced generative adversarial network can be used as a generic framework for multi-agent learning at the system level. 
\end{abstract}

}


\maketitle


%
\IEEEpeerreviewmaketitle


%
%
%
%
\IEEEraisesectionheading{\section{Introduction}\label{sec:introduction}}

\IEEEPARstart{P} ractical applications of multi-agent systems, such as autonomous airborne base stations, need accurate real-time state estimation and learning capabilities to achieve optimised trajectory planning with a maximised global goal (the maximum number of users connected). The airborne base stations have a high degree of autonomy, serving their own users within the signal coverage range. At the same time, however, multiple airborne stations need to be coordinated with each other to serve the users in a constantly changing environment: users move around, and neighbouring base stations produce signal interference. It is challenging to create optimised trajectories for all agents (base stations) at the same time, while considering the current state of neighbouring agents and their performed actions. 

Traditional centralised algorithms for multi-agent systems allowing agents to share their information with a central node are computationally expensive. Reinforcement learning-based distributed algorithms can only explicitly share information with their neighbours, which is computationally efficient compared with centralised algorithms. Reinforcement learning has had great successes in solving multi-agent collaborative tasks when three constraints are met: i) having an environment, ii) having a reward generation process either through an approximated function or a simple greedy method, and iii) having agents interacting with the environment. However, those methods often have high variance in their results. In addition, agents may start to compete instead of collaborating with each other due to scarce resources, such as communication channel capacities. To date, most solutions concentrate on the learning of individual agents, but neglect approaches at the system level \cite{ref1}.  

A conventional reward mechanism in a reinforcement learning framework is designed to produce a cost-benefit assessment of a given action, and subsequently apply a high or low reward in a heuristic search approach~\cite{ref2}. Hjeldm et al.\ proposed an approach to intelligent drone tracking using reinforcement learning~\cite{ref3}. This type of mechanism required continuous feedback from the environment, which means this technique scaled poorly to a large team with multi-agents; furthermore, traditional Bayesian approaches to reinforcement learning problems cannot model the inherent variability of the action of the state. Deep Q Network (DQN) reinforcement learning method used multiple hidden layers of a neural network to fit a state value and a state-action value distribution into a single agent~\cite{ref3}.  

The goal of classical reinforcement learning was to find an optimal policy that could maximize rewards. Instead of searching for a policy, researchers proposed a heuristic search which aimed to improve an estimated state value function, in order to minimize the expected distance between the value function's output and agents's tates~\cite{Inverse1}. Since heuristic search methods were impossible to emulate all states in an environment . Other researchers focused on finding a reward function that 'experts' are implicitly optimising either from states to actions, or from states to reward values, often called inverse reinforcement learning~\cite{Inverse2}. Two distinct approaches of inverse reinforcement learning were explored: the first method was proposed to seek a way for  directly approximating the reward function by tuning inputs~\cite{Inverse2}; the second one was focused on learning a policy that matched its action with demonstrated behaviour~\cite{Inverse3}.  The first approach depended on selecting a complete reward structure or set of feature functions which can be hard to generalise since various environments may use different reward features~\cite{Inverse4}. The second approach was sensitive for deterministic actions since the optimization became theoretically impossible if the underlying reward function was a non-convex function~\cite{Inverse5}. The inverse reinforcement learning method had generalisation issues since it used ``experts'' to demonstrate how the optimal behaviour should be, which was unrealistic in real-life.  Furthermore, agents based on inverse reinforcement learning methods were only assumed to follow an optimal policy and were prone to only take suboptimal actions by the agents.

\color{black}

By contrast, generative adversarial networks (GANs) have shown remarkable results at generating data that imitates a data distribution in a multi-agent system~\cite{ref4}. For example, an extended generative adversarial imitation learning framework was proposed in~\cite{ref5} to train an action policy network for autonomous vehicles, with only one action of the vehicle modelled. Although a large number of studies on image data augmentation tasks provided promising results, to our knowledge, far fewer research efforts have been devoted to the application of adversarial training to collaborative multi-agent systems.

Inspired by the study~\cite{ref6}, we propose a deep generative model and an opponent discriminator model followed by a two-player min-max game formula~\cite{ref3} as the single learning process for modelling the behaviour of the entire team. We built on a reward mapping method that combines adversarial generative networks with the use of reinforcement learning to produce domain-specific rewards. The generative model is used to examine the distribution of the value function for all agents, in order to reduce the training steps while optimising the overall result. As aforementioned, we applied our model to a practical case study: the optimised deployment of airborne base stations in mobile communications. In our approach, the generator is trained to generate a predicted reward map and trained adversely with  discriminator, so that the networks optimise the properties of distributed multi-agent behaviours in an adversarial fashion. Our results show that Reward-Reinforced Generative Adversarial Networks (RR-GANs) are able to achieve the best global goal values, compared to other state-of-the-art reinforcement learning methods. This is thanks to how our model can represent the distribution of the value function, replacing the approximation of Bellman updates via an adversarial learning method.


\section{Methodology}

The modern field of reinforcement learning is based on optimal control~\cite{Bellman}, which came into use during the 1960's to describe the problem of designing a controller to minimize a measurement of a dynamic system's behaviour .  The Bellman equation, as defined in~(\ref{bellman}), states that the value of the initial state must equal the value of the expected next state, plus the expected reward along the way. Instead of modelling the expectation of each value, we design a new method which applies Bellman's equation to provide an approximated optimal value function.  The distribution of $Z$ is characterized by the interaction of two distributions: the reward $R$ distribution, the probability that the tuple $(s,a)$  at time $t$ will lead to the next distribution $Z$ at time $t+1$. 
\begin{equation}
Z(s, a) \equiv^{D}  R(s,a)+Z(\acute{S},\acute{A}) 
\label{bellman}
\end{equation}

To solve the Bellman equation, we aim to recover a estimation probability distribution over the reward function from ``expert'' demonstrations using generative adversarial modelling.  At the initial stage, agents exploit the environment and use maximum likelihood rewards to choose best actions. At the execution stage, the generative modelling is used for generated reward mapping. Over time the agent is supposed to customize its actions to the environment so as to maximize the sum of this reward. 

\color{black}
\subsection{Generic GAN}

The generator $G$ uses the original environment input $o$ and a randomly selected reward experience $n$ to generate the predicted reward map $p$. Or, in mathematical notation, $G:\{o, n\} \rightarrow p$. The adversarial discriminator $D$ tries to classify the $o$ concatenating with $p$ and the $o$ concatenating with real reward map $n$.

\begin{equation}
\begin{aligned} \mathcal{L}_{GAN}(G, D)=& \mathbb{E}_{o, p}[\log D(o, p)]+\mathbb{E}_{o, n}[\log (1-
D(o, G(o, n))]\end{aligned}
\label{lgan}
\end{equation}

As shown in~(\ref{lgan}), the $G$ aims at minimizing the objective $\mathcal{L}_{GAN}$ and the $D$ tries to maximize the objective $\mathcal{L}_{GAN}$.

\begin{equation}
\mathcal{L}_{L 1}(G)=\mathbb{E}_{o, p, n}\left[\|p-G(o, n)\|_{2}\right]
\label{l1}
\end{equation}

We substitute the definition given by~(\ref{l1}) into~(\ref{gg}). The generator objective for optimising the divergence between generated data and the input real data is:
\begin{equation}
G_{global}=\arg \min _{G} \max _{D} \mathcal{L}_{G A N}(G, D)+\lambda \mathcal{L}_{L 1}(G)
\label{gg}
\end{equation}

\subsection{RR-GAN}
The data used for learning is obtained by a greedy method. An agent only takes the action when the number of connected users increases, where a reward is given. At the learning stage, the user distribution ( a part of the environment) is received as the input to the generator network: the target location is the one with the maximum reward. 

The GAN is a multi-agent system with $N$ agents and the loss function $\mathcal{L}_{GAN}(G,D)$ is denoted as $f([\theta_{1},\theta_{2},\dots,\theta_{N}], \phi)$ , where $\theta_{N}$ is the parameter vector of the $N^{th}$ agent and $\phi$ is the parameter vector of the environment (i.e. the user distribution), and where the objective functions of generator are $f(\theta_{N})$  and the objective function of discriminator  $-f(\theta_{N})$. The optimum is a Nash equilibrium defined as in (~\ref{eq3}).
\begin{equation}\label{eq3}
 \Theta \in argmax f(\Theta, \phi^{*}), \phi^{*} \in argmax-f(\Theta^{*}, \phi)
\end{equation}
  
The maximum likelihood estimator solves $\frac{d\mathcal{L}}{dx}=0$, where $x = (\Theta, \phi)^T$. The Nash-equilibrium points pose the following properties:
 
 \begin{equation}\label{key}
\frac{ d\mathcal{L}_{GAN}}{d\Theta} = 
 \begin{bmatrix}
 df(x*) \\
 -df(x*)  
 \end{bmatrix}
 =0
 \end{equation}
 
 \begin{equation}
 \label{gradient}
 T(x^{*}) = \left[\begin{matrix}\frac{d^2f\left(x^\ast\right)}{d^2\theta}&\frac{df\left(x^\ast\right)}{d\theta d\Phi}\\ \\-\frac{d^2f\left(x^\ast\right)}{d\theta d\Phi}&-\frac{d^2f\left(x\ast\right)}{d^2\mathrm{\Phi}}\\\end{matrix}\right] \leq 0
 \end{equation}
 
We substitute the definition of $T(x)$ and gradient of $\frac{d\mathcal{L}_{GAN}}{d\theta}$, as given by (\ref{key}) and~(\ref{gradient}) into~(\ref{eq:key}). Using consensus optimization~\cite{mescheder2017numerics}, the objective function of the generator can add a 2-norm as a regularization term. The updated rule for the generator network becomes:
 \begin{equation}\label{eq:key}
 x_{k+1} = x_{k} + \alpha\frac{d\mathcal{L}_{GAN}}{d\theta}+\gamma T(x)^{T}
 \end{equation}
 
When $\gamma$ and $\alpha$ are proper values using hypothesis testing, the optimisation will be locally stable at the Nash equilibrium if $T(x)$ is inevitable.  Hence, a stationary distribution exists, and the GAN architecture can converge to this stationary distribution. Furthermore, to avoid low convergence speed or even divergence, we used the regulated RMSProp optimisation~\cite{Arjovsky}. The empirical results reported in the results section demonstrate the ability of our method to solve the reinforcement learning tasks. Further mathematical proof requires an argument similar to~\cite{Arjovsky}. 

RR-GAN is used to generate the predicted reward maps from the current environment input and randomly selected past rewards from each agent. This method reduces the fitting difficulty of the network of the discriminator $D$. The network mainly needs to learn from the data source  (i.e. user distribution), and the time varying environment (i.e. how many users are within range of each airborne base station).  In the initial stage an agent is placed in a situation without knowledge of any goals or other information about other agents. As an agent acts in the environment, a reinforcement reward governs its actions: increasing the number of connected users is rewarded. By only giving the agent a reward when connected users are reached, the agent learns to achieve its goals. However, since each agent competes (due to signal interference), generative modelling is used to generate reward mapping at each execution stage. Over time, the agent customizes its actions to the environment to maximize the sum of the rewards. 

\begin{figure*}
	\centering  
	\includegraphics[width=1\linewidth]{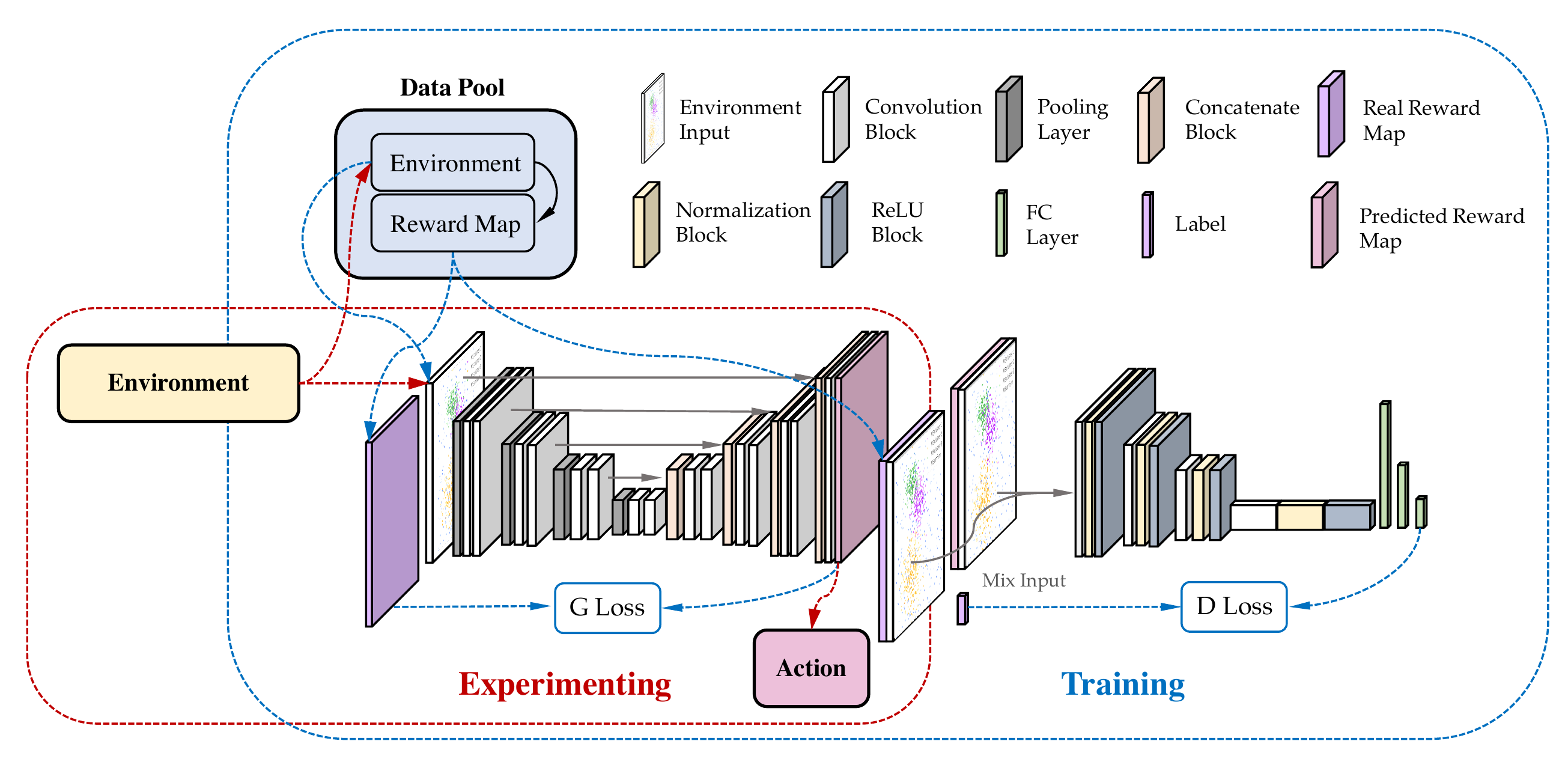}  
	\caption{General diagram of the proposed RR-GAN }  
	\label{fig:1}   
\end{figure*}

As shown in Fig.~\ref{fig:1}, real-time information from the environment (in this case, the user distribution) allows the network to self-adjust. Agents will act according to the predicted reward map for moving in the right direction. At the training stage, all agents will use a greedy method to exploit the maximum users connections in their own signal coverage areas to generate a reward map (blue arrows in Fig.~\ref{fig:1}). After several iterations, all agents adjust their behaviours according to the generated reward maps (red arrows in Fig.~\ref{fig:1}).

  \cite{pan2020loss}.
 
\noindent The pseudo code of RR-GAN is shown below:\\

\begin{algorithm}
\caption{Reward-Reinforced GAN}\label{algorithm}
\KwData{Environment $E$, Predicted Reward Map $R_{predicted}$, Reward Map $R$}
\KwResult{Trained Reward-Reinforced GAN parameters $\theta$}
	$\text {Randomly generate environment } E$\;
	$\text {Exploit rewards using greedy methods and store reward map } R$\;
	\For{every epoch}{
		$\text {Train Generator by using stored } E \text { and } R \text {, (Update } \theta \text {)}$\;
		$\text {Train Discriminator by using stored } R \text { and } R_{predicted} $\;
		\For{every iteration}{
			\If{environment changed}{
			$\text {Generator generate } R_{predicted}$\;
			$\text {Find global optimal position}$\;
			}
			$\text {Do action toward global optimal}$
		}

	}
\end{algorithm}

 Two aspects can be modified to enhance the stability of GAN training: model setup, and optimization methods. In this work we used the adversarial learned kernels in a batch training. 
 
 In our experiment, the generative neural network is composed of both the generator model and discriminator model. Our network structure follows Goodfellow's published work~\cite{0h}. The generator model consists of a U-net network, which is composed of two fully connected layers and eight deconvolutional layers. 
  The discriminator is composed of five convolution layers followed by two fully connected layers. The convolution and deconvolution layer come with a batch normalization layer.  The output layer of the generator uses the sigmoid function as the activation function while all other and discriminator network are based on ReLu as  activation functions.  Both generator and discriminator networks are trained under the Adam solver~\cite{kingma2014adam} with a learning rate of 0.0001~\cite{isola2017image, zhu2017unpaired}. The parameters are selected with reference to some other GANs~\cite{isola2017image, zhu2017unpaired}. 
 
 The specific structures of the generator and the discriminator are shown in Tables~\ref{gan_g}, \ref{gan_g_b}, and~\ref{gan_d}.
 
 \begin{table}[htbp]
 	\centering
 	\caption{Detailed model architecture ---\\Reward Map Prediction Network generator}
 	\begin{tabular}{cccc}
 		\toprule
 		\textbf{Layer name} & \textbf{Block type} & \textbf{Output resolution} & \textbf{Output depth} \\
 		\midrule
 		Down 1 & Conv Block & 100$\times$100 & 64 \\
 		Down 2 & Bottleneck & 50$\times$50 & 128 \\
 		Down 3 & Bottleneck & 25$\times$25 & 256 \\
 		Down 4 & Bottleneck & 12$\times$12 & 512 \\
 		Down 5 & Bottleneck & 6$\times$6   & 1024 \\
 		Up 1  & Bottleneck & 12$\times$12 & 512 \\
 		Up 2  & Bottleneck & 25$\times$25 & 256 \\
 		Up 3  & Bottleneck & 50$\times$50 & 128 \\
 		Up 4  & Bottleneck & 100$\times$100 & 64 \\
 		Out Conv & Conv Block & 100$\times$100 & n \\
 		\bottomrule
 	\end{tabular}%
 	\label{gan_g}%
 \end{table}%
 
 \begin{table}[htbp]
 	\centering
 	\caption{Bottleneck architecture}
 	\begin{tabular}{cccc}
 		\toprule
 		\textbf{Layer name} & \textbf{Out Direction} & \textbf{Kernel size} & \textbf{Stride size} \\
 		\midrule
 		Conv 1+BatchNorm+ReLU &       & 3$\times$3    & 1 \\
 		Conv 2+BatchNorm+ReLU & Up & 3$\times$3    & 1 \\
 		Max Pool & Down & 2$\times$2    & 2 \\
 		\bottomrule
 	\end{tabular}%
 	\label{gan_g_b}%
 \end{table}%

 \begin{table}[!htbp]
 	\centering
 	\caption{Detailed model architecture ---\\Reward Map Prediction Network discriminator}
 	\begin{tabular}{cccc}
 		\toprule
 		\textbf{Layer name} & \textbf{Kernel size} & \textbf{Output depth} & \textbf{Stride size} \\
 		\midrule
 		Conv 1+BatchNorm+ReLU & 4$\times$4    & 64    & 1 \\
 		Conv 2+BatchNorm+ReLU & 4$\times$4    & 128   & 1 \\
 		Conv 3+BatchNorm+ReLU & 4$\times$4    & 256   & 1 \\
 		Conv 4+BatchNorm+ReLU & 4$\times$4    & 512   & 1 \\
 		Conv 5+BatchNorm+ReLU & 4$\times$4    & 512   & 1 \\
 		Fully Connected 1+Dropout & —     & 128   & — \\
 		Fully Connected 2+ReLU & —     & 1     & — \\
 		\bottomrule
 	\end{tabular}%
 	\label{gan_d}%
 \end{table}%

\section{Experimental design and results}

In this section, we applied the proposed approach to control the trajectory of airborne base stations in a mobile communication system. It is worth noting that this approach can solve many other tasks; for instance, trajectory prediction for industrial robotic collaboration in warehouses~\cite{ref7}. More importantly, it can also be applied to common multi-agent problems: multi-agent learning can exhibit unexpected interactions between agents as they gravitate toward an equilibrium~\cite{ref8}.

\subsection{Problem Statement}

\begin{figure}[h]
	\centering  
	\includegraphics[width=0.8\linewidth]{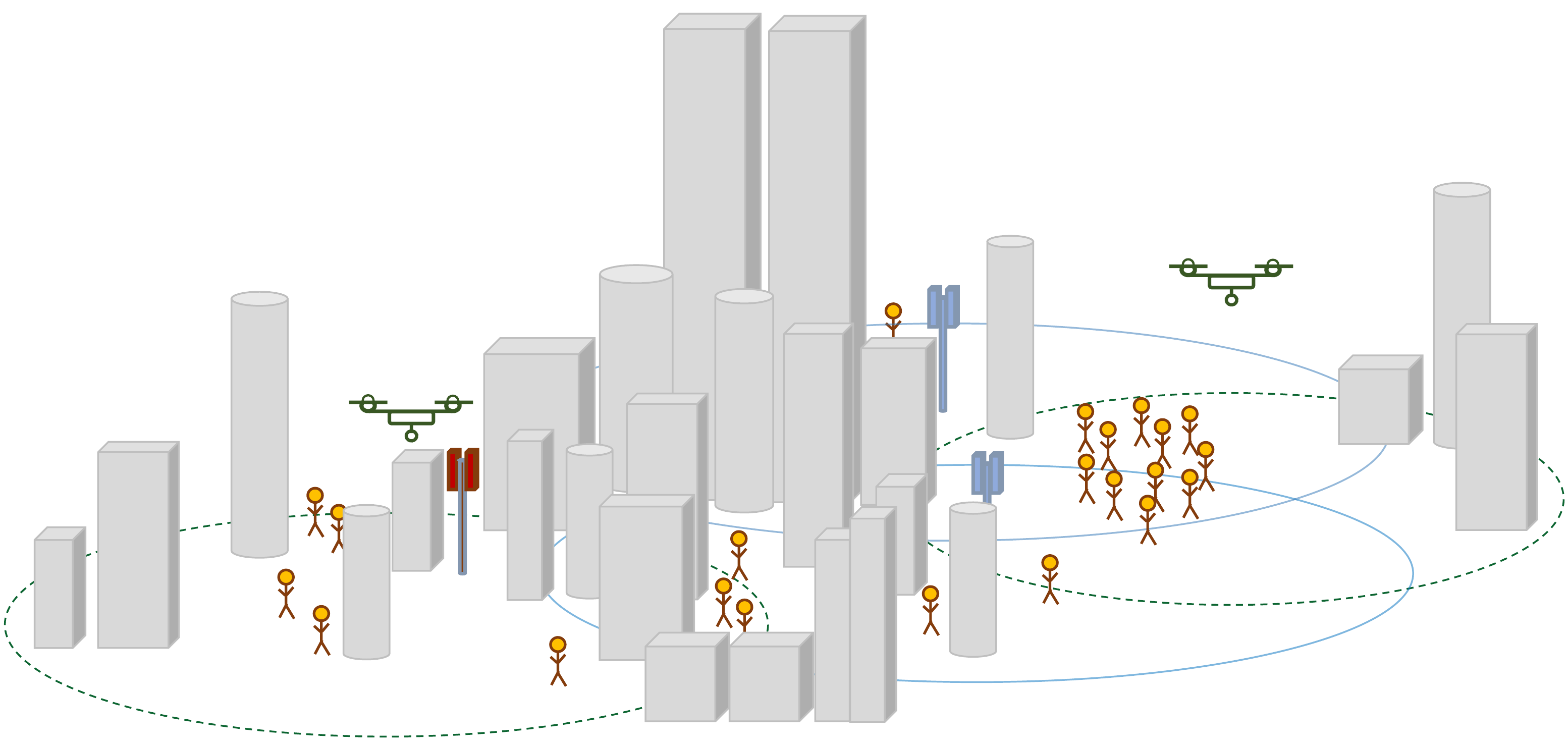}  
	\caption{Possible Airborne Base Stations using conditions: a static base station (red) has malfunctioned; people gather at an area with poor coverage from existing base stations}
	\label{fig:2} 
\end{figure}

In a mobile communication network, there are various scenarios where sudden spikes in connection demands can be generated, such as large social events (e.g. business campaigns, political rallies, university opening ceremonies or other unexpected events), or emergencies like the malfunction of the existing mobile base stations. Airborne base stations can provide a fast response to these situations, as shown in Fig.~\ref{fig:2}. It is important to precisely control the movement of the airborne base stations, in order to connect the maximum number of users to the mobile network.

While providing mobile communications with airborne base stations, interference and noise are two potential impairments to signal quality~\cite{06}. Two parameters model communication channel (signal-to-interference-plus-noise ratio (SNIR), and reference signal received power (RSRP)) can be computed as the Power Density multiplied by the Antenna Effective Area~\cite{06,07}. SNIR and RSRP are used for evaluating the potential signal quality between the mobile station and a user device. A threshold is used to distinguish the connectivity between users and base stations. The overall attenuation effects can be represented by path loss (\emph{free-space loss}). The behaviour of the airborne base stations depends on their proximity to the neighbouring airborne base stations, and how many users are being served. Our objective is to learn a joined distribution of the reward map of each airborne base station for making long-term predictions about the optimised number of users staying connected to a mobile network.

The communication model follows the formula in (\ref{loss}). All parameters in Table~\ref{environment_comm} are remained the same in all experiments.

\begin{equation}
	L_{\mathrm{s}} = \left(\frac{ 4 \pi d }{{\lambda}}\right)^2
\label{loss}
\end{equation}
 In (\ref{rsrp}), $c$ is the speed of light in metres per second; EIRP is the Equivalent Isotropic Radiated Power (the drone transmit power) in watts; $f_{\mathrm{c}}$ is the carrier frequency in hertz; $d$ is the distance between the user and the airborne base station in metres and the wavelength is $\lambda = \frac{c}{f_{\mathrm{c}}}$.
\begin{equation}
R S R P_{n, u}= \frac{EIRP}{L_{\mathrm{s}}}= \frac{EIRP\; c^{2}}{\left(4\pi  f_{\mathrm{c}} d\right)^{2}}
\label{rsrp}
\end{equation}

In (\ref{rsrp}), the RSRP for the link between the user $u$ and the airborne base station $n$ is calculated according to the EIRP, and the free space path loss is given in  (\ref{loss}).
\begin{equation}
S I N R_{n, u}=\frac{R S R P_{n, u}}{N+\sum_{\forall i \neq n} R S R P_{i, u}}
\label{sinr}
\end{equation}

The Signal to Interference plus Noise Ratio (SINR) is defined in (\ref{sinr}), where $N$ is the noise power in Watts.

\begin{table}[htbp]
  \centering
  \caption{Communication and environment parameters}
    \begin{tabular}{lr}
    \toprule
    \textbf{Parameters} & \textbf{Value} \\
    \midrule
    Communication Parameters &  \\
\cmidrule{1-1}    Lowest SINR requirment, $\theta_{\mu}$ & 0dB \\
    UAV-base station antenna directivity angle, $\phi_{a p}$ & 60$^{\circ}$ \\
    Carrier frequency, $f_{\mathrm{c}}$ & 2.4GHz \\
    Drone transmition power & 40dBm \\
    Bandwidth & 200kHz \\
    Noise power spectral density & $10^{-20.4}$W/Hz \\
    \midrule
    Environment Parameters &  \\
\cmidrule{1-1}    Random seed & 19 \\
    Length of the area & 100m \\
    Width of the area & 100m \\
    Drone step size  & 10m \\
    Cluster number & 4 \\
    Ratio of users of each cluster & 4:5:6:6  \\
    Total number of users, $N_{\mathrm{u}}$ & 1050 \\
    Drone numbers, $N_{\mathrm{d}}$ & 1,2,4,8 \\
    User Height, $h_{\mathrm{u}}$& 1.5m \\
    Drone Height, $h_{\mathrm{d}}$& 30m \\
    \bottomrule
    \end{tabular}%
  \label{environment_comm}%
\end{table}%

As an example, the distribution of users (and their mobile phones) can be modeled by a bivariate distribution consisting of a mixture of distinct Gaussian clusters\cite{0a}. The probability of users appearing can be treated as a mixture of two time-invariant 2-dimensional Gaussian distributions varying with time\cite{08,09}. Each airborne base station have many users (mobile phones)~\cite{0a}. For each Gaussian user cluster, the ``user appearing'' probability also follows a 2-dimensional Gaussian distribution. The users in each mobile communication cell follows the distribution in (\ref{pl}).

\begin{equation}
(X_{pl}, Y_{pl})\sim N\left(X_{gc}, \sigma_{gc1}^{2}, Y_{Gc}, \sigma_{gc2}^{2}, \rho_{gc}\right)
\label{pl}
\end{equation}

The distribution $f_{pl}(x, y)$ of user locations follows a 2-dimensional Gaussian distribution, where $X_{gc}, Y_{Gc}$ is the cluster center. The standard deviation, including $gc1$, $gc2$, $pl1$ and $pl2$, are random variables.

\begin{equation}
(X_{cn}, Y_{cn}) \sim N\left(X_{pl}, \sigma_{pl1}^{2}, Y_{pl}, \sigma_{pl2}^{2}, \rho_{pl}\right)
\label{ms}
\end{equation}

The user distribution $f_{cluster}(x_{cn}, y_{cn})$ for cluster $n$, in (\ref{ms}), follows a 2-dimensional Gaussian distribution~\cite{0f}.
\begin{equation}
f_{MS}(x, y) = \frac{1}{k+1}\left(uniform + \sum_{n=1}^{k}cluster_n\right)
\label{add}
\end{equation}

\begin{figure}[h]
	\centering  
	\includegraphics[width=\columnwidth]{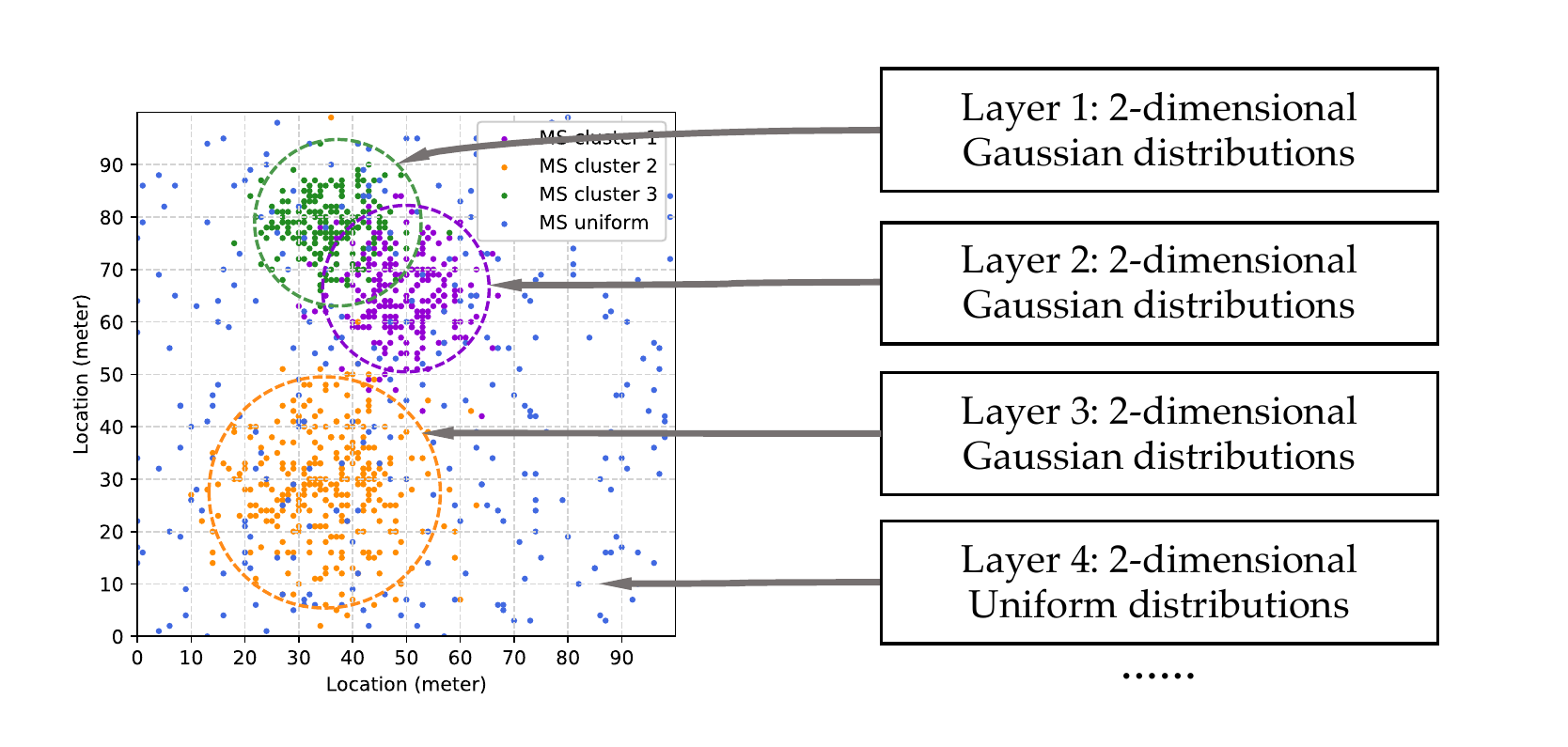}  
	\caption{User distribution}  
	\label{fig:env_cons}   
\end{figure}

The distribution of users is defined in four layers in our simulation, with three 2-dimensional Gaussian layers (layer 1, layer 2 and layer 3) and one uniform distribution layer (layer 4), as shown in Fig.~\ref{fig:env_cons}. The orange cluster in Fig.~\ref{fig:env_cons} shows the biggest size of the cluster compared with the green and purple clusters, where these three clusters simulate an emergency scenario with many users. The fourth layer is a uniform distribution layer, where the normal distribution is used to simulate a general scenario. The number of users at each Gaussian-distributed layer are 300, 250, and 200 respectively. The standard deviations of each cluster are 10, 7, and 6 (\ref{add}). 
\color{black}

\subsection{Results}

Each airborne base station chooses one action among five options at each iteration: move ``east'', ``west'', ``south'', or ``north'', or stay at the same spot. All base stations start at the right bottom corner. If an airborne base station moves in a direction which increases the number of users, the reward for this airborne base station will increase.  However, if this airborne base station is too close to the neighbouring base station, the neighbouring base station will lose users. The global goal is to provide connectivity for the maximum number of users.

We compare the performance of the proposed method with the following baseline models: Q-learning, SARSA, DQN and k-means. All the baseline models use the same input features, and are trained with the same iterations and the same learning rate. 
\color{black}
\begin{itemize}
	
	\item	Q-learning is a model-free off-policy algorithm \cite{02}, which provides agents with the capability of learning to act optimally in Markovian domains \cite{01}. The reward map is exploited and updated through the Q-table~\cite{0e}. Fig.~\ref{fig:QLearning} shows the dispersion of the global rewards over each training episode in the airborne base station system using the Q-learning algorithm. As the graph describes, the global rewards stabilize around values between 250 and 300, with a median of 269 connected users. The maximum value of the global reward is not within the first and third quartiles.
	\begin{figure}
       \centering
        \includegraphics[width=0.4\textwidth]{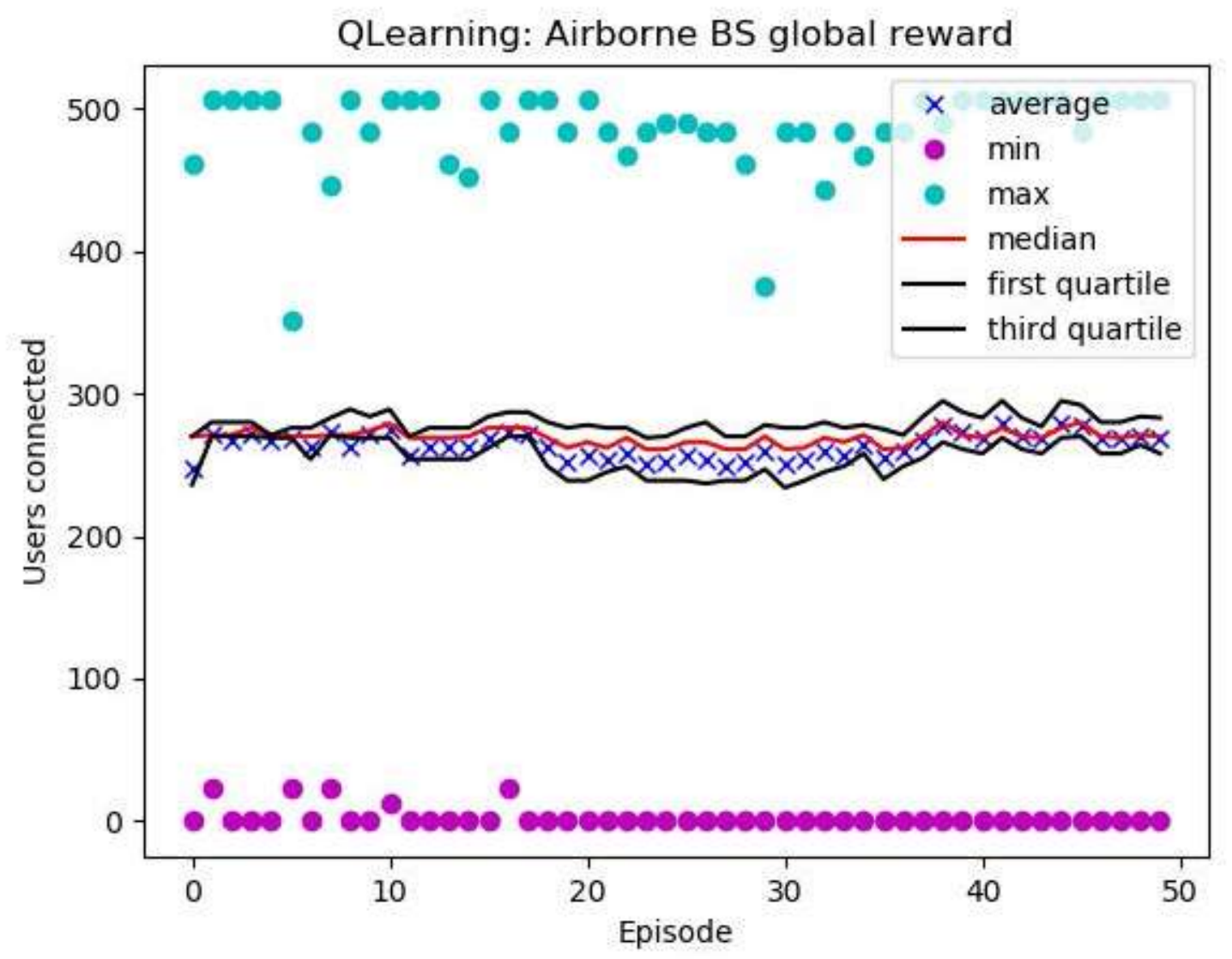}
        \caption{Reward analysis of Q-learning}
        \label{fig:QLearning}
        \vspace{-10pt}
    \end{figure}
	
	\item	SARSA applies the same policy in both data generation and evaluation. SARSA follows a Markovian (history-independent) structure \cite{0c,0d,0e}. Under this structure, the value function can be expressed in an algorithmic form known as the Bellman equation, and used to update Q-values in the lookup tables. 
	
	\item A Deep-Q Network (DQN) can directly use deep learning methods to bridge the divide between high-dimensional inputs and agent actions\cite{dqn}. The Deep Neural Network predicts the Q value of the current or potential states and actions. Two convolutional neural networks constantly update its parameters to learn the optimal option.
	
	\item A clustering network divides data objects with high similarity into clusters \cite{kmeans}. Currently popular clustering methods are k-means and mean-shift. In this work, the k-means algorithm is used as one of the  baseline methods. It is interesting to test whether a clustering simulation of the user distribution favours the clustering method. The position of the ABS is randomly chosen, and their moving direction is controlled using L2 distance followed by~(\ref{L2distance}).
	\begin{equation}
    a_{j}=\frac{1}{\left|c_{i}\right|} \sum_{x \in c_{i}} x
    \label{L2distance}
     \end{equation}
	The positions of the ABSs $a_{j}$ are updated with the number of users $x$ in each center.  As shown in Fig.~\ref{fig:fig4reward}, the k-means method actually performed very poorly, especially in the scenarios with 4 and 8 airborne base stations.
	
    \begin{figure*}[]
    	\centering
    	\includegraphics[width=.8\textwidth]{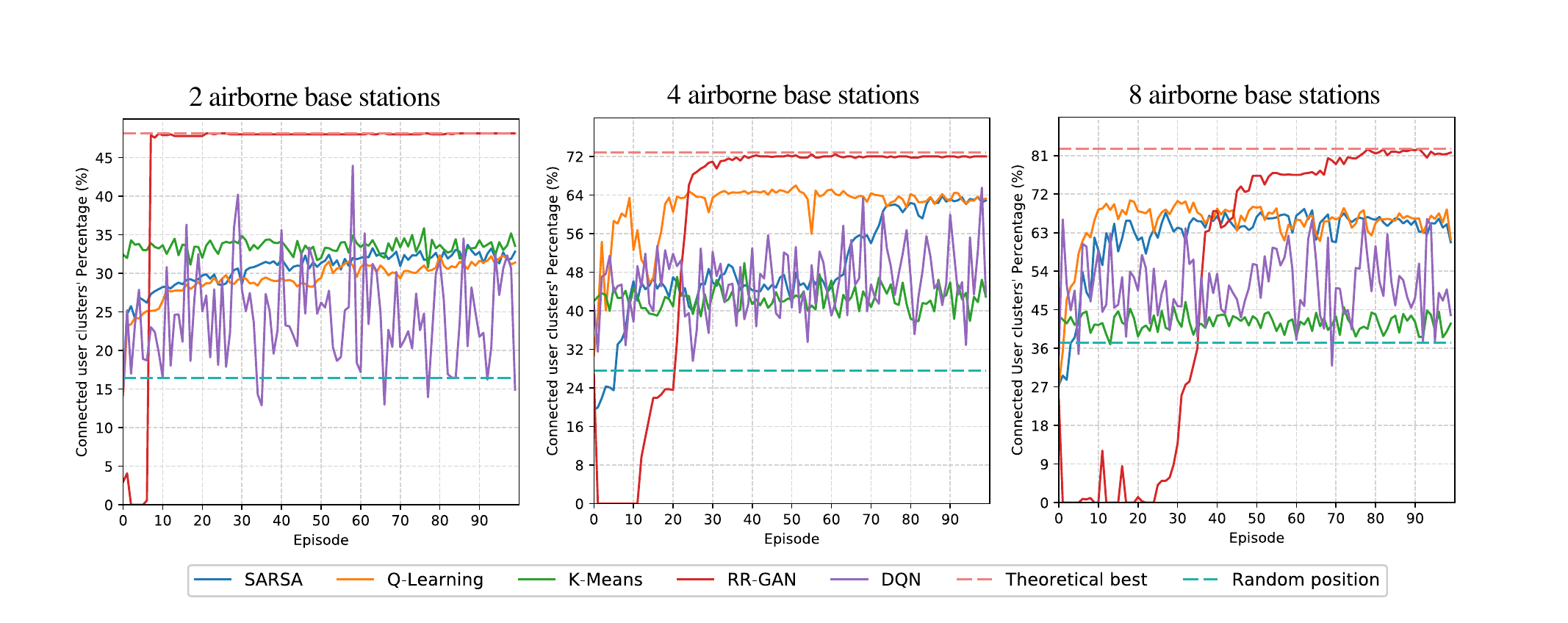}
    	\caption{Comparison among Q-learning, SARSA, k-means, DQN and RR-GAN}
      	\label{fig:fig4reward}
    \end{figure*}
	
\end{itemize}

 Fig. \ref{fig:fig4reward} compares the performance of RR-GAN against the four other baseline methods. As illustrated in Fig.~\ref{fig:fig4reward}, the RR-GAN method attains the highest average percentage of connected users over all airborne base stations, even with an increased number of airborne base stations. Since other methods do not consider the positions of the other base stations and the uncertainty during the initial learning period, RR-GAN can achieve the highest number of user connections.The theoretical best performance is calculated using a heuristic method. The performance of random position experiments is simulated when Airborne base stations moves following a random walk distribution.  It is worth noting that the DQN method cannot minimize the divergence of the generated sample distribution and over-fits the training set by maximizing the likelihood, which reduces model generalisability. 
\color{black}

\subsection{Robustness Comparison }

The potential application environments for airborne base stations are complex, real-world domains, which need to overcome the problems of high sample complexity and brittle convergence properties, requiring less meticulous hyperparameter tuning. Therefore, here different learning rates, greedy factors and user distribution are all tested; to allow the ABS to be fully aware of the neighbouring ABS, we also tested on 100 rounds, 1,000 rounds, 10,000 rounds and 100,000 rounds. The reward map is generated based on various amounts of exploration, stopping if there is no higher total reward for the last $n$ rounds of stochastic exploring. The different number of $n$ rounds is used to test network performance. Fig~\ref{fig:fig7robustness} shows a testing environment with 2 ABSs (airborne base station), 4 ABSs, and 8 ABSs.

For all rounds, the time taken to reach maximum performance follows the same trend, particularly so for 8 airborne base stations. For 2 or 4 ABSs, a shorter exploring time may introduce oscillations in performance, but all achieve a similar plateau within a longer time frame. 

A comparison of robustness with other baseline algorithms can be found in Fig.~\ref{fig:fig8robustness}. The greedy parameter is a key parameter in evaluating the algorithm’s performance to value their new actions. During the changing of the value of greedy parameters, we can investigate how each method responds with learning information and timing needed for training. With all else held equal, the greedy policy is tested under the values 0.7, 0.8, and 0.9, with the learning rate and discount factor fixed to 0.1 and 0.5, respectively. The user environment is generated with the same random seed.

As shown in Fig.~\ref{fig:fig8robustness}, final performance improves as the learning rate approaches 1. Additionally, the larger the greedy factor is, the faster the algorithm will converge. Although initial efficacy is low, RR-GAN has a steeper convergence rate when compared to other baseline methods and, most importantly, is the only method to reach the global optimal. It is noted the DQN method is not chosen in the greedy method comparison. The DQN method hasn't performed well since state transition probability is approximated with one hidden layer neural network. 

\subsection{Scalability Test}

To investigate the impact of the number of user clusters on the performance of  RR-GAN, we ran a scalability test to investigate how much the number of user clusters inevitably affected the RR-GAN method. As shown in Fig.~\ref{fig:fig9robustness}, even with an increased number of user clusters, RR-GAN performed consistently well with 2 airborne base stations, 4 airborne base stations and 8 airborne base stations.   
\subsection{Learning from the Neighbouring Base Station}

\begin{figure}[h]
	\centering  
	\includegraphics[width=1.0\linewidth]{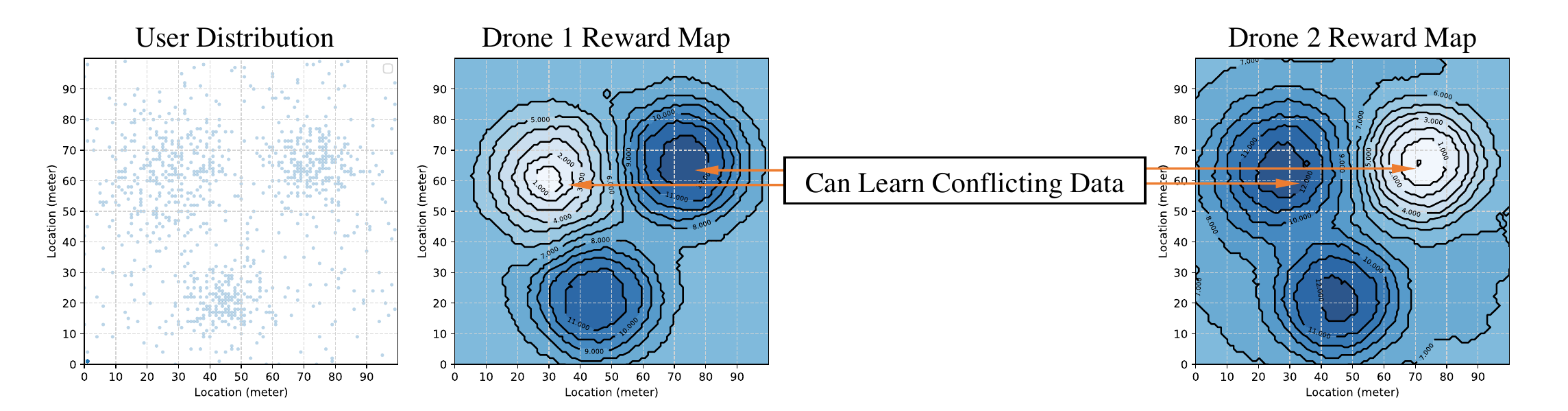}  
	\caption{Learning from neighbour}  
	\label{fig:Conflict_Data}   
\end{figure}

Airborne base stations will interfere with each other if they are too close. When airborne base stations move close to a user cluster, the reward increases; however, if a neighbouring airborne base station has already moved towards that centre, the reward decreases. This experiment uses an environment of two airborne base stations to demonstrate that our RR-GAN can be made aware of the neighbouring base station via checking the history rewards of neighbouring ABSs. As shown in Fig.~\ref{fig:Conflict_Data}, the first channel of the reward map is generated for the first ABS. This airborne base station gains awareness of the fact that the left user cluster has the second optimal reward. For that reason, the network predicts that the second airborne base station will move to that place in next epoch, which results in the first ABS not moving into that direction. Similarly, the second ABS, with the help of the second channel of the reward map, will predict and find the global optimal for the first ABS and automatically reduce the rewards. In this manner, RR-GAN can successfully predict the neighbouring ABS trajectory to achieve the best global goal. 
\begin{figure*}[h!]
	\centering
	\includegraphics[width=.7\textwidth]{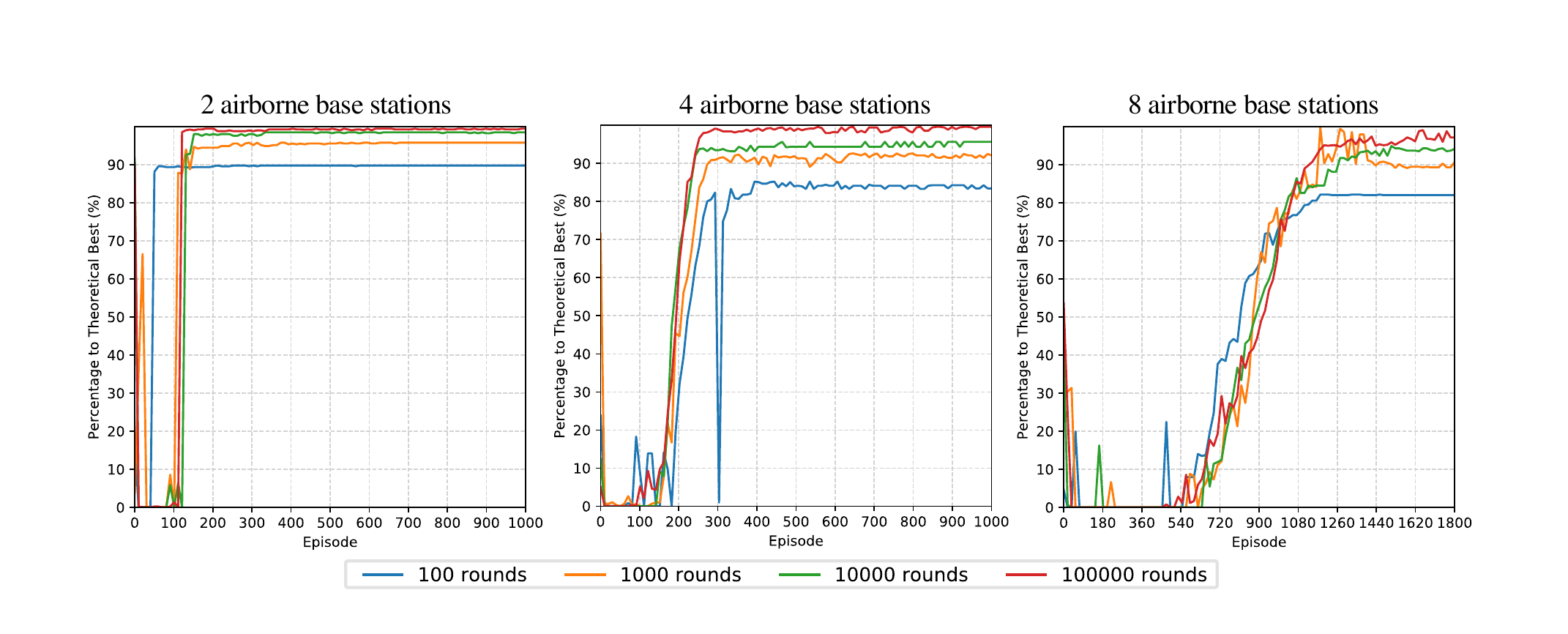}
	\caption{Robustness comparison with various greedy methods}
	\label{fig:fig7robustness}
\end{figure*}

\begin{figure*}[h!]
	\centering
	\includegraphics[width=.7\textwidth]{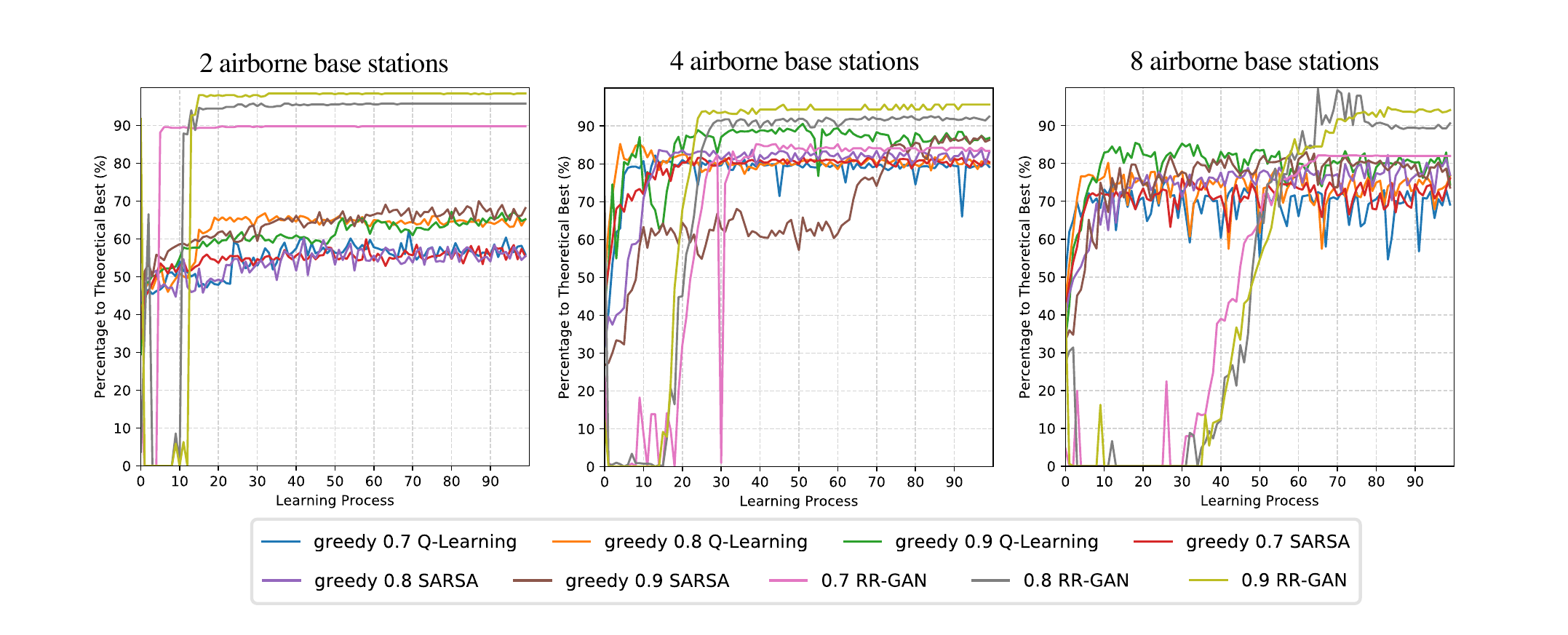}
	\caption{Robustness comparison among Q-learning, SARSA, K-means, DQN and RR-GAN with various greedy methods}
	\label{fig:fig8robustness}
\end{figure*}

\begin{figure*}[h!]
	\centering
	\includegraphics[width=.8\textwidth]{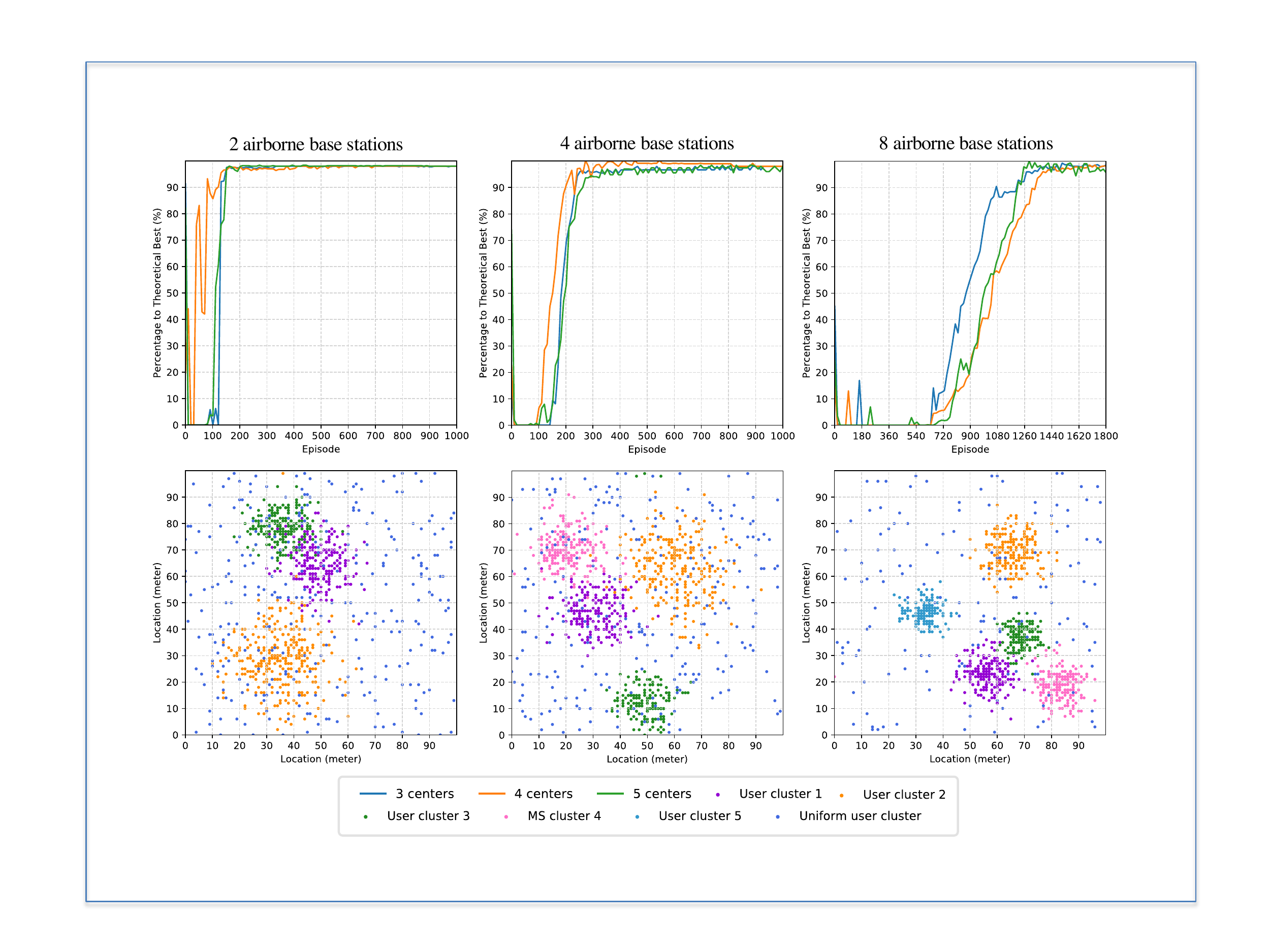}
	\caption{Robustness comparison among environment with different amount center}
	\label{fig:fig9robustness}
\end{figure*}

\section{Conclusion and Discussion}

In this paper, a Reward-Reinforced GAN (RR-GAN)-based generic framework for multi-agent systems is proposed, which are potentially generalised to any multi-agent system. We propose to use a generator network as an implicit indication of the distribution of users, in order to achieve the maximum global goal. Our case study demonstrates how RR-GAN has the best system performance compared with other baseline methods (the source code can be accessed via GitHub link: https:$\backslash$$\backslash$github.com$\backslash$Changgang-Zheng$\backslash$Reward-reinforced-Generative-Adversarial-Network $\backslash$). In terms of future developments, while this work demonstrated the advantages of using RR-GAN for multi-agent learning, several challenges still remain:

\begin{itemize}
	
	  \item \textbf{Convergence Failures:} The convergence properties for generative neural networks constitute an open research question. The theoretical condition for the adversarial model is when the loss function is a class of convex optimisation algorithms ~\cite{pan2020loss}. In that case, the adversarial model can be guaranteed to find a unique solution. However, when neural networks are used for the generator and the discriminator (as in this paper), it is not always guaranteed to converge to a unique solution. We used Nash equilibriums from game theory to demonstrate the possibility of converging for an adversarial model when a small change in probabilities for discriminator leads to a situation where two conditions hold: the generator did not change and has no better strategy in the new environment. At the training stage, it is a challenge to modify GAN design so that the discriminator can be trained optimally, avoiding the issue of convergence failures. The current research is investigate how to move away from unimodal distributions as a natural relaxation to solve potential convergence failures~\cite{GAN52}.
	\color{black}
	
	\item \textbf{Uncertainty Quantification: } Accurately estimating user movements is commonly based on Bayesian methods, which introduces epistemic uncertainty into the model parameters (cluster centres and random seeds)~\cite{van2020dependent}. Gaussian processes present issues of model inadequacy and parameter uncertainty when scaling to high-dimensional problems. Our RR-GAN network creates out-of-distribution samples so that classifiers can be explicitly taught about uncertain inputs. We evidence the complications in the training process from computer simulations, the implications of which for real-life experimental data are still unknown.
	
	\item \textbf{Explanation of uncertainty: } Creating explanations for the causes of model uncertainty \cite{yang2017single} is a relatively under-explored area. In a GAN, uncertainty may arise because an input is unlike the training data and has been constrained by a set of known features in a previous unseen combination. This type of explanation has only been very recently explored by Merric and Taly to calculate the variance of Shapley values \cite{merrick2019explanation}, despite the fact that it remains a challenging research area which may produce important consequences for assessing multi-agent learning systems. 
	 
	 \item \textbf{Interpretability:} Although we demonstrated that RR-GAN works as a stochastic estimation to improve system performance, there remain issues with interpretability. One solution is to use models that are intrinsically interpretable, such as logistic regression, so that an accurate explanation can be produced. However, the increasing states of multi-agent will create the issue of retrieving historical samples: consequently, in the future we will investigate attempts to provide solutions for checking the state of every timestamp: for instance, using queries on temporal graphs when dealing with increasing large state spaces~\cite{parra2020temporal}.
\end{itemize}

 \section*{Acknowledgements}

We would like offer our sincerest gratitude to Paulo Valente Klaine and Jo\~{a}o P.B. Nadas, who
provided an initial discussion for this project. 
\newpage

\bibliographystyle{IEEEtran}
\bibliography{ref.bib}




\end{document}